\documentstyle[aps,prl,epsfig]{revtex}
\begin{document}
\newcommand{\identity}{\bf 1}
\newcommand{\ket}[1]{| #1 \rangle}
\newcommand{\bra}[1]{\langle #1 |}
\newcommand{\braket}[2]{\langle #1 | #2 \rangle}
\newcommand{\tr}{\rm tr}
\newcommand{\rank}{\rm rank}
\newcommand{\proj}[1]{| #1\rangle\!\langle #1 |}
\newcommand{\ba}{\begin{array}}
\newcommand{\ea}{\end{array}}
\newtheorem{theo}{Theorem}
\newtheorem{defi}{Definition}
\newtheorem{lem}{Lemma}
\newtheorem{exam}{Example}
\newtheorem{prop}{Property}
\newtheorem{coro}{Corollary}

\twocolumn[\hsize\textwidth\columnwidth\hsize\csname
@twocolumnfalse\endcsname

\author{Peter W. Shor$^*$, John A. Smolin$^\dag$ and
Ashish V. Thapliyal$^\ddag$}

\title{Superactivation of Bound Entanglement}

\address{\vspace*{1.2ex} \hspace*{0.5ex}{$^*$AT\&T Labs--Research,
Florham Park, NJ 07932, $^\dag$IBM T.J. Watson Research
Center, Yorktown Heights, NY 10598, $^\ddag$University of
California at Santa Barbara, Department of Physics, Santa Barbara,
CA 93106\\}}

\date{May 26, 2000}

\maketitle
\begin{abstract}
We show that, in a multi-party setting, two non-distillable
(bound-entangled) states tensored together can make a distillable
state.  This is an example of true superadditivity of distillable
entanglement.  We also show that unlockable bound-entangled states
cannot be asymptotically unentangled, providing the first proof that
some states are truly bound-entangled in the sense of being both
non-distillable and non-separable asymptotically.

\end{abstract}

% close bracket for pretty quant-ph mode
]
%
%\narrowtext

The joint state of more than one quantum system cannot always be
thought of as a separate state of each system, nor even as
a correlated mixture of separate states of each system
\cite{epr}, a situation known as quantum entanglement.  Entanglement
leads to the most counterintuitive effects in quantum mechanics,
including the disturbing idea due to Bell that quantum mechanics is
incompatible with local hidden variable theories \cite{bell}.  Even
today new quantum oddities with their basis in entanglement are being
found, and the study of entanglement is at the heart of quantum
information theory.

A state belonging to parties $A$, $B$, $C$, {\em etc.} is said to be
{\em inseparable} if it cannot be written in separable form
\begin{equation}
\rho^{ABC\ldots} =\sum_i p_i \rho_i^A \otimes \rho_i^B \otimes \rho_i^C \ldots
\end{equation}
for any positive probabilities $p_i$ summing to one and set of density
matrices $\rho_i^A,\rho_i^B,\rho_i^C \ldots$, where, for example,
$\rho_i^A$ operates on the Hilbert space belonging to party $A$.
%and in general may not be the same as $\rho_i^B$.
Notice that the superscripts
$A$, $B$, $C$, {\em etc.} denote the parties by whom the state is shared.
%In the rest of this letter we will follow this convention.
We say that a state is {\em distillable} if some pure entangled
state shared by some subset of the parties is obtainable
(asymptotically \cite{ashishandjohn}) from it by local operations
and classical communication (LOCC) amongst the parties.

It is known that many inseparable quantum mixed states are
distillable, while separable states are not \cite{purification,bdsw}.
More recently it has been shown that some mixed states which are entangled
in the sense of being inseparable nevertheless cannot be distilled into any pure
entanglement \cite{horodeckibound1,horodeckibound2}.  Such states are
known as {\em bound-entangled} states.

It has been an open question whether bound-entangled states, though
inseparable, are actually entangled at all in an asymptotic sense.
A state $\rho$ is said to be {\em asymptotically unentangled} \cite{ashishandjohn})
if for any positive $\epsilon$ there exists a
number of copies $N$, a number $m$ sublinear in $N$ of EPR pairs
shared in some way among the parties, and an LOCC method of constructing
from those EPR pairs a state $\rho'$ such that
$F(\rho^{\otimes N},\rho')> 1-\epsilon$
for some sensible definition of the fidelity $F$ between
two density matrices \cite{asymptotic-separability}.
In this letter we show the first example of a bound entangled state that
can be proved not to be asymptotically unentangled.  Other examples can
be found it \cite{VC}.

In the bipartite case, bound entanglement may sometimes be useful in a
kind of quasi-distillation process known as {\em activating} the bound
entanglement \cite{activation} in which a finite number of
free-entangled mixed states are distilled with the help of a large
number of bound-entangled states.  This is not a true distillation of
the bound entanglement in that no more pure entanglement is produced
than the distillable entanglement of the free-entangled mixed states,
the distillable entanglement being defined as the pure entanglement
distillable per state from an infinite number of copies of a state.

In the case of more than two parties the bound entanglement can be
more truly activated by the presence of free entanglement.  In
examples given by Cirac, Tarrach and D\"ur \cite{dur1,dur2,dur3}, and
in the equivalent formulation of unlockable bound-entangled states
\cite{quant-ph/0001001}, when several parties share certain bound
entangled states, and when some subset of the parties get to share
pure entanglement then some pure entanglement may be distilled between
parties where it would be impossible to obtain any without having
shared the bound-entangled state.  This is a kind of superadditivity
of distillable entanglement, though in the known cases no {\em more}
entanglement is distilled than the pure entanglement that was shared,
rather it is in a different place.  Later in this letter we will look
at unlockable states in much more detail since we will need some of
the results about them.

In this letter we present an effect we call superactivation of bound entanglement.
It is ``super'' in the sense of {\em superadditivity} of distillable entanglement,
but without the restrictions of either of the earlier types of activation of
bound entanglement.  In superactivation two entangled mixed states $\rho,\rho'$
are combined to yield more pure entanglement than the sum of what a set of parties
could distill from either $\rho$ or $\rho'$ on their own, even if many copies
of $\rho$ or $\rho'$ are shared.  In particular, both states in our example
are bound-entangled states from which no pure entanglement can be distilled.
Our result thus provides the first example of superaddivity of distillable
entanglement.

We will use the usual notation for the maximally entangled states
of two qubits (the Bell states):
\begin{equation}
\ket{\Psi^\pm}=\frac{1}{\sqrt{2}}\left(\ket{\!\uparrow\downarrow}
\pm\ket{\!\downarrow\uparrow}\right),
\ \ket{\Phi^\pm}=\frac{1}{\sqrt{2}}\left(\ket{\!\uparrow\uparrow}
\pm\ket{\!\downarrow\downarrow}\right)
\end{equation}
For convenience we adopt the following notation as well:
\begin{eqnarray}
\label{bellstates}&\Psi=\{\Psi^-,\Psi^+,\Phi^+,\Phi^-\} {\rm\ with\ elements\ }
\Psi_i{\rm\ , and}\\
&\sigma=\{\identity_2,{1\ \ 0 \choose 0\ -1}, {0\ -1 \choose 1\ \
0}, {0\ 1 \choose 1\ 0}\} {\rm\ with\ elements\ } \sigma_i
\enspace ,
\end{eqnarray}
where $\identity_2$ is the identity matrix in $2\times 2$. In the
text, we shall refer to a Bell state as any one of the four states
(\ref{bellstates}) and to an EPR state as the standard singlet
state $\ket{\Psi^-}$. The Bell states $\ket{\Psi_i}$ are related
to the standard EPR state $\ket{\Psi^-}$ by the following
identities, up to an overall phase which is unimportant here:
\begin{eqnarray}
\label{tostandard}&&\ket{\Psi^-}= \identity_2 \otimes \sigma_i
\ket{\Psi_i} = \sigma_i \otimes \identity_2 \ket{\Psi_i}\\
\label{psii}&&\ket{\Psi_i}= \identity_2 \otimes \sigma_i
\ket{\Psi^-} = \sigma_i \otimes \identity_2 \ket{\Psi^-}\enspace.
\end{eqnarray}
In teleportation \cite{teleportation}, $A$ and $B$ share an EPR
pair $\ket{\Psi^-}$, and $A$ has another qubit in a state
$\ket{\psi}$. $A$ first does a joint measurement on her two qubits
in the basis formed by the Bell states. There are four equally
likely outcomes corresponding to the Bell states $\ket{\Psi_i}$.
$B$'s half of the EPR pair after this measurement is $\sigma_i
\ket{\psi}$ up to a phase that can be ignored. Then $A$
communicates $i$ to $B$ who then performs a rotation $\sigma_i$ on
his state giving $\sigma_i^2 \ket{\psi}$. But $\sigma_i^2 = \pm
\identity_2$ and thus the final state $B$ has is $\ket{\psi}$ up
to a phase.

An easy lemma about teleportation is that if a state $\ket{\psi}$
is teleported from $A$ to $B$ using an incorrect one of the Bell
states $\ket{\Psi_i}$ rather than $\ket{\Psi^-}$ as normally
required by the protocol, then the result of the teleportation
will be $\sigma_i \ket{\psi}$, again up to an overall phase.  This
is easily seen by using (\ref{psii}) to write the incorrect Bell
state as $\ket{\Psi^-}$ with a $\sigma_i$ operating on $B$'s part
of the $\ket{\Psi^-}$ i.e. $\ket{\psi} = \identity_2 \otimes
\sigma_i \ket{\Psi^-}$. If A's outcome of the Bell measurement is
$j$ then B's corresponding state is $\sigma_i \sigma_j
\ket{\psi}$.  Thus after B applies the rotation $\sigma_j$ the
state becomes $\sigma_j \sigma_i \sigma_j \ket{\psi}$.  If the
rotation $\sigma_j$ which is the final step in teleportation could
be squeezed in before the $\sigma_i$ the proof would be complete,
but instead it follows the $\sigma_i$. However, the rotations used
in teleportation are also the $\sigma$ matrices, and all the
$\sigma_i,\sigma_j$ either commute or anticommute
($\sigma_i\sigma_j =\pm \sigma_j\sigma_i$) and so their order can
be freely interchanged up to a phase.  Thus the lemma is proved.
$\Box$

In \cite{quant-ph/0001001} a four-party bound entangled state
was presented:
\begin{equation}
\rho^{ABCD}=\frac{1}{4} \sum_{i=0}^3 \ket{\Psi_i}^{AB}\bra{\Psi_i}
\otimes \ket{\Psi_i}^{CD}\bra{\Psi_i}
\label{thestate}
\end{equation}
%\begin{eqnarray}
%\nonumber &&\rho^{ABCD}=\frac{1}{4} \left(
%\ket{\Phi^+}_{AB}\bra{\Phi^+}\otimes\ket{\Phi^+}_{CD}\bra{\Phi^+}\right.\\
%\nonumber &&\ \ \ \ \ \ \ \ \ \left.+\right.\ket{\Phi^-}_{AB}
%\bra{\Phi^-}\otimes\ket{\Phi^-}_{CD}\bra{\Phi^-}\\
%&&\ \ \ \ \ \ \ \ \ \ \left.+\right.\ket{\Psi^+}_{AB}\bra{\Psi^+}\otimes\ket{\Psi^+}_{CD}
%\bra{\Psi^+}\\
%\nonumber &&\ \ \ \ \ \ \ \ \ \ \ \left.+\right.\ket{\Psi^-}_{AB}
%\bra{\Psi^-}\otimes\ket{\Psi^-}_{CD}\bra{\Psi^-}\left.\right)
%\label{thestate}
%\end{eqnarray}
In other words, $A$ and $B$ share one of the four Bell states, but don't
know which one, and $C$ and $D$ share the same Bell state, also
not knowing which one.

This state has several properties:

%\begin{itemize}
\noindent $\bullet$
%\item
Symmetry under interchange of parties:
$\rho^{ABCD}=\rho^{ABDC}=\rho^{ADBC}$, {\em etc.}  This may be
verified by writing out the state as a $16\otimes 16$ matrix and
interchanging indices.  A more enlightening way is to use our lemma and
think of the state in terms of teleportation.  First, we note that
some of the symmetries are obvious, for example interchanging $A$ and
$B$ because Bell states are themselves symmetric under interchange.
So the only symmetry we need to consider is the interchange of
$B$ with $C$ and the rest can be constructed trivially.

Consider the state in its original form, with $A$ and $B$ sharing
an unknown Bell state and $C$ and $D$ sharing the same one.  Now
consider $A$ and $C$ getting together and performing a Bell
measurement and obtaining the result $\ket{\Psi_j}$, which we can
think of as $A$ and $C$ doing the first step required to teleport
$A$'s particle to $D$ using the unknown Bell state shared by $C$
and $D$.  The result $\ket{\Psi_j}$ is random since $A$ and $C$
had halves of completely separate unknown Bell states.  The state
being teleported is half of a Bell state given by Eq.\
(\ref{psii}) $\sigma_i \otimes \identity_2 \ket{\Psi^-}$ as is the
state used in the teleportation.  So, by our lemma, if the
teleportation were completed an extra $\sigma_i$ would be
introduced, and the two $\sigma_i$'s would cancel being
self-inverse (up to a phase).  Thus, $B$ and $D$ would share a
standard $\ket{\Psi^-}$.  But if the $\sigma_i$ needed to complete
teleportation is not performed, this means that $B$ and $D$ share
the Bell state $\sigma_j^{-1} \otimes \identity_2 \ket{\Psi^-} =
\sigma_j \otimes \identity_2 \ket{\Psi^-}=\ket{\Psi_j}$ (ignoring
phases), which is the result obtained by $A$ and $C$.  So $AC$ and
$BD$ share identical random Bell states, which was the original
form of the density matrix, but with $A$ and $C$ interchanged.

\noindent $\bullet$
%\item
Non-distillability: When all four parties remain separated and
cannot perform joint quantum operations, then they cannot distill any
pure entanglement by LOCC,
even if they share many states, each having density matrix
$\rho^{ABCD}$.  This comes from the fact every party is separated from
every other across a separable cut.  This is easy to see since the
state (\ref{thestate}) is separable across the $AB:CD$ cut by
construction and the state has the symmetry property.

\noindent $\bullet$
%\item
Unlockability: The entanglement of the state can be
{\em unlocked}.  If $A$ and $B$ come together and perform a joint
quantum measurement, they can determine which of the four Bell states
they have (the four Bell states form an orthogonal basis) and tell $C$
and $D$ the outcome.  Since $C$ and $D$ then know which Bell state
they share, they can convert it into the standard $\ket{\Psi^-}$ state
using local operations by Eq.\ (\ref{tostandard}).  Because of the
symmetry property any two parties can join together to help the other
two get a $\ket{\Psi^-}$.  Note that the unlockability property
implies the state must not be fully separable, or no entanglement
could be distilled between separated parties, even when some of the
parties come together.
%\end{itemize}
Because the state is both non-distillable and entangled, it is by
definition a {\em bound-entangled state}
\cite{horodeckibound1,horodeckibound2}.

\begin{figure}
\epsfxsize=7cm
\epsfbox{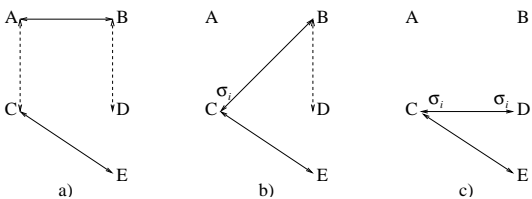}
\caption{How to distill the state $M$ into an EPR pair between $D$ and
$E$: a) The state $M$, with the two identical but unknown Bell states
of $\rho^{ABCD}$ shown as dashed arrows, and those of $\rho^{ABCE}$ as
solid arrows.  b) $A$ has teleported her half of the unknown Bell
state she shares with $B$ to $C$, using the unknown Bell state
$\ket{\Psi_i}$ she shares with $C$.  The state has picked up a factor
of $\sigma_i$.  c) $B$ has teleported his half of the unknown Bell
state he shared originally with $A$ and now shares with $C$ to $D$
using his unknown Bell state shared with $D$ (again, $\ket{\Psi_i}$).
The state has picked up another factor of $\sigma_i$.  The $\sigma_i$'s
cancel each other and the final state of $CDE$ is of the form of
Eq. (\protect\ref{thestate}), but with party $C$ having two of the
qubits {\em i.e.} $\rho^{CCDE}$.  This is the unlockable bound-entangled
state of \protect\cite{quant-ph/0001001} in its ``unlocked'' configuration,
and can therefore be distilled into a $DE$ EPR pair by $C$ simply
measuring which Bell state she has and telling $D$ and $E$
which one they have since the two are the same.
}
\label{figure}
\end{figure}

Now we consider the mixed state of five parties $A$, $B$, $C$, $D$, and $E$
\begin{equation}
M=\rho^{ACBD}\otimes\rho^{ABCE}
\label{thesuperactivatedstate}
\end{equation}
where $\rho^{ACBD}$ (call it state 1) and $\rho^{ABCE}$ (call it state 2)
are the states of
Eq.\ (\ref{thestate}) but with the qubits assigned to different
parties.  Thus parties $A$, $B$, $C$ and $D$ each have a one qubit subsystem
of state $\rho^{ACBD}$ and similarly parties $A$, $B$, $C$ and $E$ each
have a one qubit subsystem of state $\rho^{ABCE}$.  Thus the parties $A$, $B$,
$C$, $D$ and $E$ have Hilbert spaces of size $4$, $4$, $4$, $2$, and $2$.
Technically $\rho^{ACBD}$ could be written as
$\rho^{ABCD}$ due to the symmetry property but it will be
useful to have it explicitly written in the form where it
is an unknown Bell state shared between $A$ and $C$, and
the same state shared by $B$ and $D$.  The state $M$
is illustrated in Figure \ref{figure}a.
$M$ is the tensor product of two density matrices,
neither of which is independently distillable.  We now show
how to distill a $\ket{\Psi^-}$ between $D$ and $E$.

In the distillation procedure $A$ and $B$ use state 1 to ``teleport''
state 2 to $C$ and $D$.  First, party $A$ teleports her half of the
unknown Bell state she
shares with $B$ (which is part of state 2 and shown by the solid arrow
 connecting $A$ and $B$ in Figure \ref{figure}a and part of state 2)
to $C$ using the unknown Bell state she shares with $C$
(which is part of state 1, shown by the dashed arrow connecting $A$ and $C$ in
the figure).  This results
in the situation of Figure \ref{figure}b, where now $C$ shares an
unknown Bell state with $B$, her half of which has additionally picked
up the unknown rotation $\sigma_i$ from having been teleported with an
incorrect Bell state $\ket{\Psi_i}$. The Bell state connecting $A$ and
$C$ is gone in the figure, since it has been expended performing the
teleportation.  Then $B$ teleports his half of that state to $D$ using
the unknown Bell state (again $\ket{\Psi_i}$ that they share,
resulting in the situation of Figure \ref{figure}c, where now $C$ and
$D$ share the unknown Bell state originally shared by $A$ and $B$,
both halves of which having been rotated by $\sigma_i$.  It is
important to note here that because of the structure of $\rho^{ACBD}$
this is the {\em same} $\sigma_i$.  Now, using Eq. (\ref{psii}) and
the fact that $\sigma_i^2$ is the identity (once again except for a
phase), we can see that the $\sigma_i$'s cancel and we are left with
the state $\rho^{CDCE}$.  This is the same form as the four-party unlockable
state (Eq. (\ref{thestate})) but with one party sharing two of the qubits,
and it is therefore distillable into a pure EPR pair shared by $D$ and $E$.

$M$ cannot be distilled into EPR pairs between any of the
other parties.  This is because if we give the five parties
the additional power of having $D$ and $E$ in the same room,
then $M$ is just two copies of $\rho^{ABCD}$ which are known
not to be distillable (by definition if $\rho$ is not distillable,
then neither is $\rho^{\otimes N}$).  To construct a state out
of tensor products of bound-entangled states that is distillable
into any kind of pure entanglement, it is sufficient to symmetrize
$M$, {\em i.e.}
\begin{eqnarray}
&&M_{\rm S}=\\
\nonumber&&\ \ \ \ \ \ \rho^{ABCD}\otimes\rho^{ABCE}\otimes\rho^{ABDE}
\otimes\rho^{ACDE}\otimes\rho^{BCDE}\ .
\end{eqnarray}
Then the distillation protocol just described can be used
to obtain an EPR pair between any two of the parties, and
using more copies of $M_{\rm S}$ one can obtain
EPR pairs between all pairs of parties.  Once this is accomplished
any arbitrary multi-party entangled state can be constructed by
one party creating it in his lab and teleporting the pieces
as needed to the others.

Because $M$ (Eq. (\ref{thesuperactivatedstate})) is distillable, it
cannot be that the original state $\rho^{ABCD}$ is asymptotically
unentangled.  If it were, then many copies $N$ of $\rho^{ABCD}$ and
$\rho^{ABCE}$ could be created arbitrarily precisely using a number of
EPR pairs sublinear in $N$. These could be used to create $N$ copies of
$M$ which could then be distilled into $N$ pure EPR pairs between $D$
and $E$.  These $DE$ EPR pairs would, to arbitrarily high
probability, pass any test that pure EPR pairs would pass.  Thus, an
amount of entanglement sublinear in $N$ would have been converted into
$N$ EPR pairs by LOCC, which is impossible \cite{bdsw}.

In fact, all unlockable bound-entangled states are asymptotically
inseparable.  This is because when some subset $S$ of the parties
possessing such a state come together in the same lab the state
becomes distillable.  If the state were asymptotically unentangled then
it could be made arbitrarily precisely with asymptotically no
entanglement even when parties in $S$ are actually together in the
same lab (it cannot hurt for them to be together as they can
conveniently ignore this fact as they carry out whatever procedure
results in the creation of the state).  But then they can distill a
finite amount of arbitrarily pure entanglement per state from the
sublinear amount of entanglement they started with, which is
impossible.  It is worth noting then that the unlockable
bound-entangled states are the first states shown to be true
bound-entangled states in the sense of both being non-distillable and
being non-separable asymptotically.

It is clearly a necessary condition for superactivation that at least
one of the states involved must not be asymptotically unentangled.
It is by no means a sufficient one, however, since the states
$\rho^{ABCD}$ and $\rho^{EFGH}$ are each not asymptotically unentangled
but $\rho^{ABCD}\otimes\rho^{EFGH}$ is not distillable as the
two pieces are on disconnected sets of parties.

In the individual states $\rho^{ABCD}$ and $\rho^{ABCE}$, every party
is separated from every other party by at least one separable cut.  In
order for the combined state $M$ to be distillable into a $DE$ EPR
pair, and for $M_{\rm S}$ to be distillable into EPR pairs
between any pair of parties, it is necessary that the parties who get
EPR pairs no longer be separated by any separable cut, as is indeed
the case by construction for these states.
% \cite{construction}.
Using
this observation, D\"ur has reported a whole family of superactivated
states \cite{durprivate} based on the unlockable bound-entangled
states of \cite{dur1,dur2,dur3}.  References \cite{dur1,dur2} also discuss
separable cuts and their relation to distillibility in more detail.

In conclusion, we have shown that asymptotically entangled states
exist from which no pure entanglement can be distilled.  This has
been suspected for some time but ours is the first such example
for which it has been proved.  Further we have shown the
surprising fact that distillable entanglement is not additive by
showing two undistillable asymptotically entangled states that
when combined gives a distillable state.

Many future directions are suggested by this work.  Here we have
shown four party example of a state that is asymptotically
entangled but not distillable.  An interesting question is whether
such a state can be found for two parties.  Since the first
writing of this letter, this question has been answered in the
affirmative in \cite{VC}. Another direction for further research
is to find bipartite states that show the non-additivity of
distillable entanglement.  Such examples have been shown to exist
if the NPT-boud entangled states are truly bound entangled
\cite{nptbound}.

\end{document}